\documentclass[conference]{IEEEtran}
\usepackage{cite}
\usepackage{amsmath,amssymb,amsfonts}
\usepackage{algorithmic}
\usepackage{graphicx}
\usepackage{textcomp}
\usepackage{xcolor}
\usepackage[hyphens]{url}
\usepackage{hyperref}

\def\BibTeX{{\rm B\kern-.05em{\sc i\kern-.025em b}\kern-.08em
    T\kern-.1667em\lower.7ex\hbox{E}\kern-.125emX}}

\title{\huge Splitwiser: Efficient LM inference with constrained resources}

\author{
{Asad Aali$^*$}\\
 \emph{asadaali@stanford.edu}\\
 Stanford University
\and
{Adney Cardoza$^*$}\\
\emph{adney11@cs.utexas.edu}\\
UT Austin
\and
{Melissa Capo$^*$}\\
\emph{melissayang@utexas.edu}\\
UT Austin
}

\begin{document}
\maketitle
\def\thefootnote{$^*$}\footnotetext{These authors contributed equally to this work.}\def\thefootnote{\arabic{footnote}}
\thispagestyle{plain}
\pagestyle{plain}


\begin{abstract}

Efficient inference of LLMs remains a crucial challenge, with two main phases: a compute-intensive prompt computation and a memory-intensive token generation. Despite existing batching and scheduling techniques, token generation phases fail to fully utilize compute resources, especially when compared to prompt computation phases. To address these challenges, we propose Splitwiser, a methodology that splits the two phases of an LLM inference request onto the same GPU, thereby reducing overhead and improving memory access and cache utilization. By eliminating the need to transfer data across devices, Splitwiser aims to minimize network-related overheads. In this report, we describe the basic structure of our proposed pipeline while sharing preliminary results and analysis. We implement our proposed multiprocessing design on two widely-used and independent LLM architectures: Huggingface and vLLM. We open-source our code for the respective implementations: 
\begin{enumerate}
    \item Huggingface: \href{https://github.com/asad-aali/splitwiser}{github.com/asad-aali/splitwiser}
    \item vLLM: \href{https://github.com/adney11/vllm-sysml}{github.com/adney11/vllm-sysml}
\end{enumerate}

\end{abstract}

\begin{figure*}[h]
    \centering
    \includegraphics[width=1\linewidth]{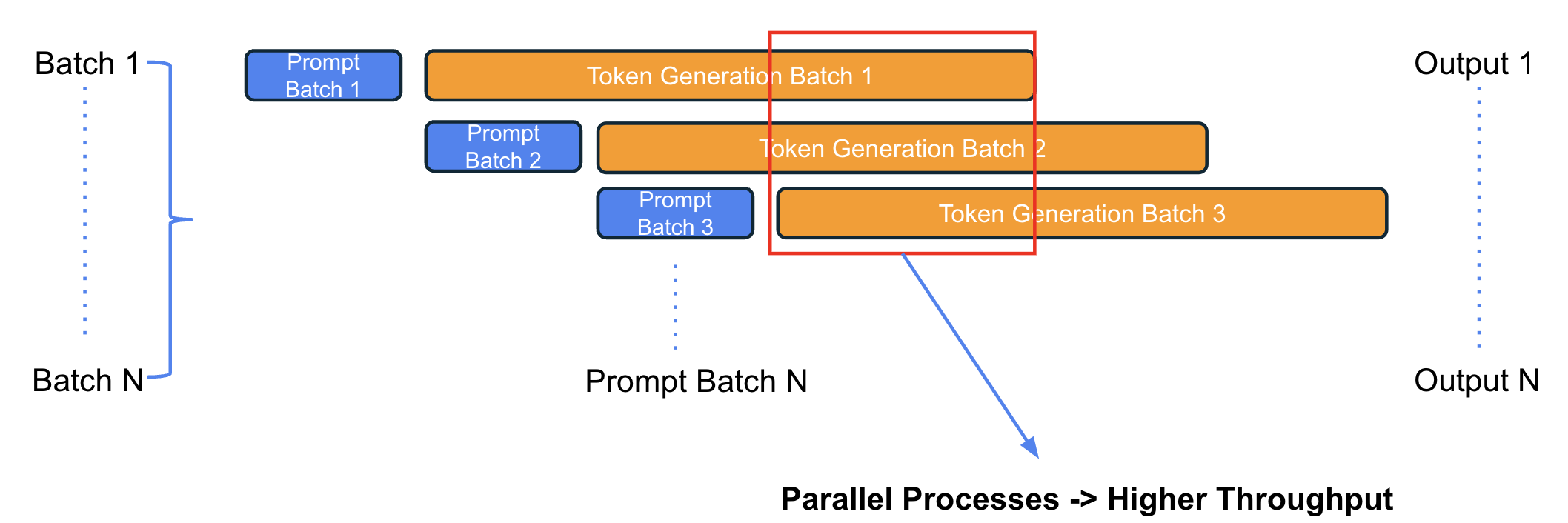}
    \caption{Splitwiser design for running prompt processing and token generation in parallel, for achieving inference speedup on a single GPU.}
    \label{fig:splitwiser_design}
\end{figure*}

\section{Introduction}

Generative Large Language Models (LLMs) have become essential in computing, offering vast capabilities in natural language processing. However, their widespread adoption has led to challenges, particularly in inference efficiency. Existing LLMs, often running on expensive GPUs, face issues with overhead and resource utilization. Prior work, specifically Splitwise \cite{patel2023splitwise}, addresses this issue by proposing to split the inference phase into two individual processes that can be run on separate GPUs. This approach allows operators to make use of the vast array of heterogeneous GPUs available in their cluster simultaneously, increasing the total hardware utilization while reducing the time to serve a batch of inference requests. The problem this work is aimed at solving is towards making split inference more accessible to operators that don't have immediate access to a vast number of GPUs, and would still like to enjoy the benefits of split inference. To that end, Splitwiser aims to address this challenge by optimizing the processing of split phase inference of LLMs on a single GPU.

\section{Background \& Related Works}
\subsection{Large Language Models}

Modern Large Language Models (LLMs) are mainly based on transformer architectures, which use attention mechanisms and multi-layer perceptron layers for input and output generation. These models, whether encoder-only, decoder-only, or encoder-decoder, have seen significant improvements in response quality and accuracy. However, their large size and complexity make operating expensive and require a lot of power.

\subsection{Generative LLM inference phases}
Generative LLM inference involves two main phases: prompt computation and token generation. During prompt computation, all input tokens are computed in parallel to generate the first token. Subsequently, the context generated from this phase is saved in a key-value (KV) cache for future token generation iterations. Token generation relies on the last generated token and the KV-cache, making its memory stronger than prompt computation.

\subsection{Batching of requests}
Batching of requests can significantly improve throughput in LLM inference. Common batching mechanisms include request-level batching, continuous batching \cite{orca-cb}, and mixed batching \cite{agrawal2023sarathi}.

\subsection{Model Parallelism}
Model parallelism, often used in training, is also applicable to inference. Splitwiser leverages model parallelism within a single GPU, dividing the workload between the prompt computation and token generation phases to improve efficiency. 

\subsection{Performance Metrics for LLMs}

Performance metrics for LLMs include end-to-end (E2E) latency, time to first token (TTFT), throughput, and time between tokens (TBT). These metrics are important to consider depending on the task, with batch tasks prioritizing throughput and latency-sensitive tasks focusing on TTFT and TBT.

\subsection{GPU clusters and interconnect}
With the increasing use of LLMs, GPU clusters have become common in cloud service providers (CSPs). These clusters typically consist of machines with multiple GPUs connected by high-bandwidth interconnects like Mellanox InfiniBand. 

\subsection{NVIDIA Multi-Process Service (MPS)}
GPUs are an essential and much-sought-out resource for inference tasks, making them quite hard to acquire from publicly available cloud service providers.
That coupled with their expensive price tags, researchers having a limited set of GPUs will not be able to run the split inference methodology to enjoy its benefits of pipeline parallelism and lower inference times.
To that end, we aim to use NVIDIA's Multi-Process Service to run multiple instances of the LLM inference serving program, to split the inference phases, on a singular GPU. Here we outline some important considerations related to MPS:
\begin{itemize}
    \item \textbf{What is MPS?} Multi-Process Service is an alternative binary-compatible implementation of the CUDA Application Programming Interface (API). It allows multiple applications to use the GPU resources simultaneously. The key benefit of MPS is that it requires no application modification, except unless there is an interdependence between the simultaneously run applications. As stated in future sections, our suggested solution does include interdependence between the applications being run using MPS, which we will discuss in detail later in this document.
    \item \textbf{How do we use MPS?} Using MPS is fairly straightforward. One has to first start the MPS daemon provided by NVIDIA, after which, simply instantiating multiple CUDA application instances that use the GPU is required. MPS manages the instances sharing the GPU resources transparently, providing faster multi-process inference times.
    \item \textbf{Challenges with MPS:} As mentioned earlier, MPS allows for multiple applications to use GPU resources simultaneously. For independent applications, this method is straightforward and requires no additional application modification. As for our case, there is a need to consume the KV-Cache data generated by the first phase, in the second phase.  In other words, application 2 will need to have a mapping between the inference request whose prompt phase is being processed in application 1, and the generated KV-cache file. Application 1 is tasked with just generating and saving the KV-Cache data, along with updating the required mapping.
\end{itemize}

\subsection{Related Work: ZeRO-Inference}
ZeRO-Inference \cite{aminabadi2022deepspeed} is a method that uses heterogeneous memory to process large language model inference using a single GPU, essentially democratizing LLM inference. By leveraging CPU memory, ZeRO-Inference enables inference of massive LLMs on a few GPUs, making massive model inference accessible.
\section{Proposed Design \& Methodolgy}
Our proposed solution is to simply start two LLM inference serving instances while using MPS.
This way, a batch of $n * 2$ inference requests could be split such that each instance individually manages the servicing of $n$ inference requests.
In the future, we will parameterize the two, so that we can serve a bigger batch of requests.\\

\textbf{What are the overheads it brings to the table?}
There are two methods here that we would like to discuss:
\begin{enumerate}
\item Running the provided vLLM engine code unmodified seems like the naive solution. However this comes with the overhead of both instances instantiating the entire model in the GPU memory.
This is a waste of GPU memory due to duplication.
\item Modifying vLLM application, so that it can use a shared memory where the model weights are loaded, alleviating the duplication problem mentioned in the above point.
This method comes with its own set of overheads. Specifically, the need to modify the vLLM code so that its workers understand how to use the shared memory, instead of allocating new memory for their model. 
Further, needing to somehow map or synchronize the two instances in a way that is throughput-efficient and correct.
\end{enumerate}

\subsection{Profiling LLM Phases}
\subsubsection{Experimental Setup}
Profiling experiments were run on an Oracle VM with NVIDIA A10 GPUs, executing single or batch LLAMA2-7B inferences with vLLM Python scripts. Model accuracy is not important for results, rather more importance is placed in controlling the number of input/output tokens and inferences batched. Inference prompts were generated with a user-specified number of random tokens and ran with a specified number of output tokens and batch size. 

\subsubsection{Measurement Methodology}
Two profiling methods were explored: NVIDIA Nsight Compute, specifically the command line version \textit{ncu}, and vLLM's internally reported metrics. SM and device memory utilization in terms of throughput can be measured with \textit{ncu} using the \textit{sm\_\_throughput.avg.pct\_of\_peak\_sustained\_elapsed} and \textit{dram\_\_throughput.avg.pct\_of\_peak\_sustained\_elapsed} metrics. We modified vLLM internal metric collection to enable "starting" and "stopping" collection rather than the default behavior of periodic reporting, allowing us to distinguish metrics collected between the prompt and token generation phases. vLLM reports device memory utilization in terms of KV cache usage \% along with other useful metrics such as average end-to-end request time and token throughput.

\begin{figure*}[h]
    \centering
    \includegraphics[width=1\linewidth]{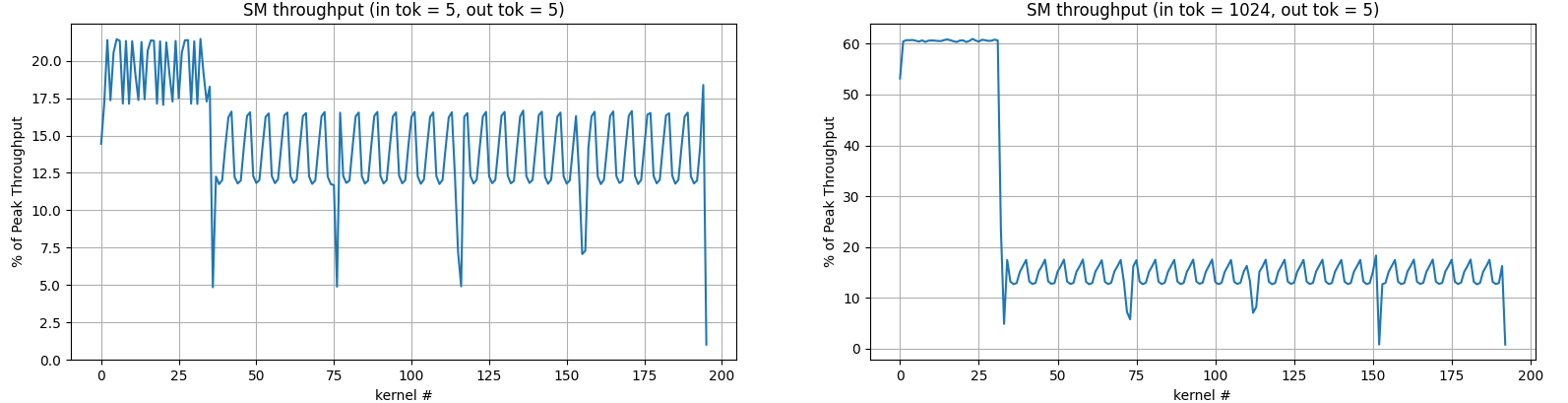}
    \caption{SM throughput comparison with an increasing number of input tokens demonstrates compute intensity in the prompt processing phase.}
    \label{fig:sm_throughput_single}
\end{figure*}
\begin{figure*}[h]
    \centering
    \includegraphics[width=1\linewidth]{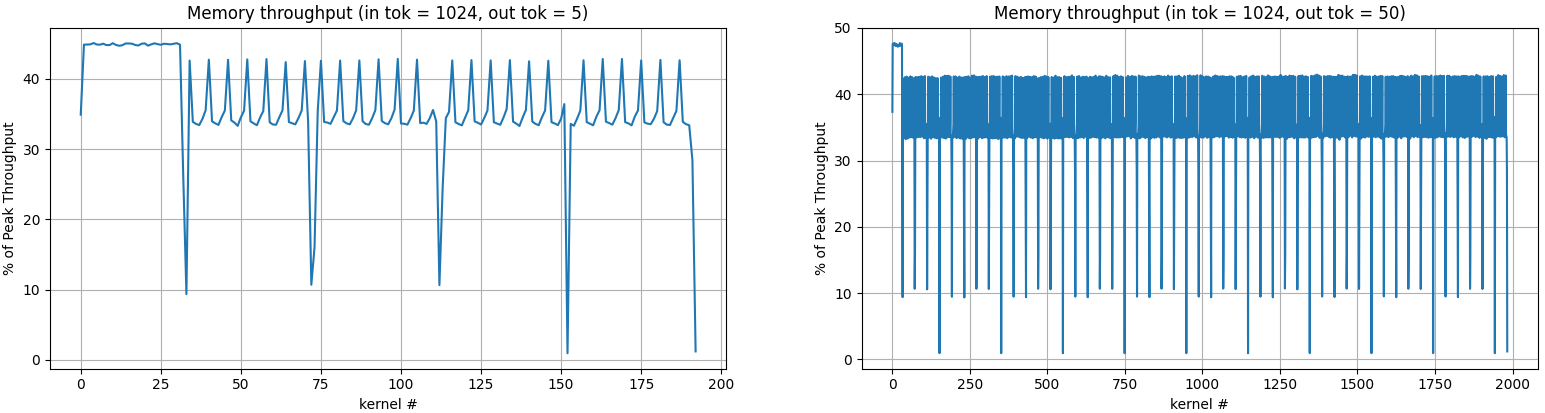}
    \caption{Memory throughput doesn't change, but increasing output tokens results in sustained memory usage.}
    \label{fig:mem_throughput_single}
\end{figure*}
\begin{figure*}[h]
    \centering
    \includegraphics[width=1\linewidth]{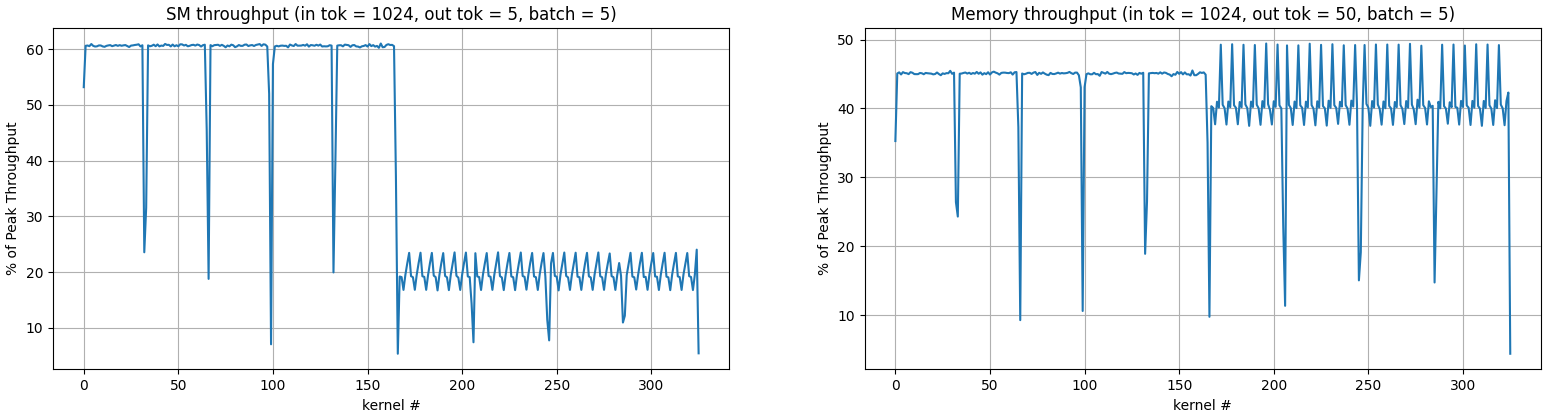}
    \caption{SM and memory throughputs running a batch of 5 inference requests. Throughputs are similar to single inference and are not fully utilized.}
    \label{fig:batch_throughputs}
\end{figure*}

\subsection{Multiprocessing/MPS with Hugginface LLMs}
\label{sec:huggingface}

We also ran experiments using the Huggingface architecture for LLMs. The goal of these experiments was to achieve inference speedup compared to the state-of-the-art performance provided by existing Huggingface implementation. In Huggingface, the inference is divided into two major phases, stemming from the transformers library:
\begin{enumerate}
    \item \textbf{Prompt Processing}: This involves the use of the Tokenizer class for prompt preprocessing to generate the encoded input. Specifically, we use the \textit{AutoTokenizer} class.
    \item \textbf{Token Generation}: This step involves calling the LLM model class itself. In our case, we use \textit{AutoModelForCausalLM} class, which accepts the tokenized prompts as input to generate the output tokens.
\end{enumerate}
We use these classes to download and load pre-trained tokenizers and models. In our case, we use the OPT-125m model. This model is relatively lightweight, making it easier to run multiple instances in parallel without overloading the GPU memory. Our experiments here follow the configurations described in Table \ref{setup_table}. Our metric of success in these experiments is improvement in latency or throughput, and we ignore the accuracy of the generated output. The input data we use includes Radiology (CT and MR) reports from MIMIC-III \cite{johnson2016mimic}. We use a pre-processed version of the data which contains 30,000 de-identified and publicly available medical records. Following the initial setup, we implement three independent inference methods:
\begin{enumerate}
    \item \textbf{Sequential Inference}: Here we use the existing Huggingface implementation without any modifications. All prompts are processed by the tokenizer. The model then takes as input the pre-processed prompts in batches to generate the final output tokens.
    \item \textbf{Splitwiser Inference}: In this experiment, we split the inference into two phases: prompt processing and token generation. Furthermore, we also split the input dataset, and create smaller sub-datasets. The number of sub-datasets is matched with the number of parallel processes running on the GPU. Next, we use PyTorch Multi-Processing to run the prompt and token phases in parallel, i.e. as the prompt processing of the first process finishes, the second process begins its prompt processing in parallel while the first process starts generating tokens, and so on. This is further described in Figure \ref{fig:splitwiser_design} 
    \item \textbf{Splitwiser + MPS Inference}: This experiment uses the same implementation as step 2), with the addition of NVIDIA Multi-Process Service (MPS), which is used for running multiple processes on the GPU.
\end{enumerate}

\subsection{Multiprocessing/MPS with vLLMs}
vLLM is a Python library for LLM inference servicing shown to greatly uplift throughput compared to standard Hugging Face pipelines through efficient memory management with PagedAttention \cite{kwon2023pagedattention} and other scheduler techniques like batching and distributed compute. The library implicitly separates prompt and token phases in the scheduler, so we are also interested to see if multiprocessing and MPS can achieve further GPU utilization in a mature scheduler implementation.

\subsubsection{vLLM Scheduler Analysis}
The scheduler in vLLM is executed at a step granularity where each step corresponds to a single token generation for a batch of inference requests. At each step, the scheduler makes a binary decision to run a batch of requests in the "waiting queue", meaning their prompts need to be processed, or to select a batch of requests in the "running queue", meaning they are in the middle of token generation. If the prompt phase batch is scheduled, the vLLM engine will then fetch and process the input tokens per request in the batch. If the token generation phase batch is selected, then the corresponding KV caches are fetched and processed. The remaining steps in pre-process, compute, and post-process the steps are identical regardless of the phase selected as they are to generate the next token. These steps are to merge the requests' inputs into a single set of inputs tensors, process these merged tensors through the LLM Sampler to generate the next token, and finally separate the Sampler output per request. Before the next scheduler step is invoked, it will determine whether any requests are completed or if any new requests were received.

It can be observed that vLLM currently implements \textbf{continuous batching}, which is where multiple requests are batched to execute simultaneously, but token and prompt phases are not. Therefore, an approach that conducts \textbf{mixed batching} where both token and prompt phases are scheduled and executed simultaneously would be the next step to improve scheduler throughput. Additionally, it should be noted that vLLM's batching implementation already increases the compute per kernel by merging a batch of requests into a single set of input tensors. Therefore, the key takeaways of the scheduler analysis are to explore mixed batching while also leveraging vLLM's batching strategy of merging tensors.

\subsubsection{Attempt 1: vLLM + Multiprocessing}
We begin with a coarse-grained approach of running two instances of vLLM on separate processes, each receiving and processing its own set of inference requests. The goal is to let the GPU hardware scheduler manage the execution of these two processes such that there is a possible overlap of prompt and token phases between the two vLLM instances. To maximize memory reuse, we needed to instantiate a single model that could be shared. The vLLM engine was modified to consume this shared model instead of instantiating individual copies on startup. For this experiment, we tested both with multiprocessing (MPx2) and with MPS (MPSx2). MPSx2 is the same as MPx2 except running as an MPS client.

\subsubsection{Attempt 2: vLLM Scheduler + Multiprocessing}
It is noted that the first attempt has scalability limitations as the number of processes and max batch size is limited by available GPU memory. As the vLLM scheduler is already capable of efficient memory management and knowing best when swapping is necessary, it would be reasonable to keep it as the singular top-level scheduler that is aware of all request contexts. So, in the following attempt, the goal is to create and execute separate processes within the scheduler explicitly when both a prompt and token phases can be scheduled.
\clearpage
\section{Experiments \& Evaluation}

The setup used to evaluate these approaches is described in Table \ref{setup_table}
\begin{table}[ht]
\caption{Experiment setup configurations}
\label{setup_table}
\centering
  \begin{tabular}{@{} l*{6}{c} @{}}
    \hline
     & \textbf{Hugging Face}   & \textbf{vLLM}    \\
    \hline
    Model    & \textit{facebook/opt-125m} & \textit{facebook/opt-125m}  \\
    GPU & A10, A100 & A10 \\
    Input Token Size & 512 & 1024 \\
    Output Token Size & 20 & 1024 \\
    Batch Size & 20 & 10, 20, 40, 80, 160 \\
    \hline
  \end{tabular}
\end{table}

\subsection{Profiling LLM Phases}
\begin{figure}[h]
    \centering
    \includegraphics[width=1\linewidth]{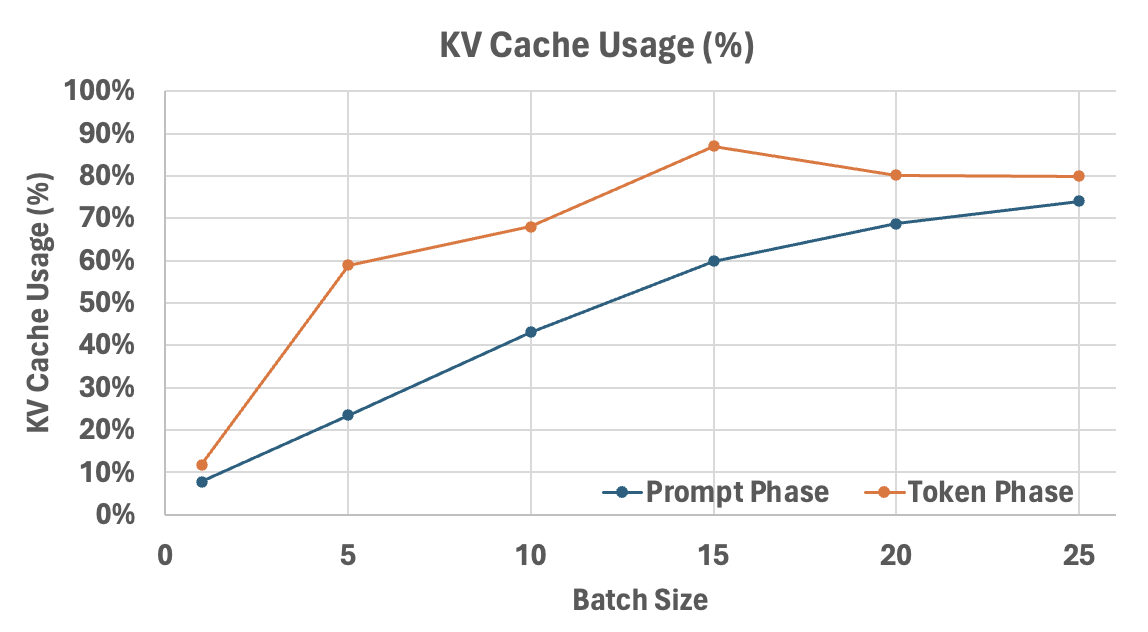}
    \caption{KV-Cache usage (\%) for a range of batch sizes, for the prompt and token phase.}
    \label{fig:kv-batch}
\end{figure}
In Figure \ref{fig:sm_throughput_single} it is observed that increasing the number of input tokens increases SM throughput across the prompt processing kernels, supporting that the prompt phase is compute intensive as the input tokens get processed in parallel.

On the other hand, from Figure \ref{fig:mem_throughput_single} while throughput doesn't increase with increasing output tokens memory, it is sustained over a longer period as tokens are generated sequentially. Noticeably, the majority of inference is spent in the token generation phase.

Batching inferences is the baseline approach to increase specifically memory usage (seen in Figure \ref{fig:kv-batch}) by parallel processing token generation for multiple requests. As expected, the usage increases as batch size increases. The trend is more stable in the prompt phase, whereas in the token phase, the trend seems to be more volatile. However, in Figure \ref{fig:batch_throughputs} it is observed that both SM and memory throughputs are comparable to the single inference scenario and not fully utilized still. Thus, the project will explore how to better push the under-utilized throughputs. Additional results from our profiling experiments are displayed in section \ref{sec:appendix} at the end of our manuscript.

\subsection{Multiprocessing/MPS with Huggingface LLMs}
\begin{figure}[h]
    \centering
    \includegraphics[width=1\linewidth]{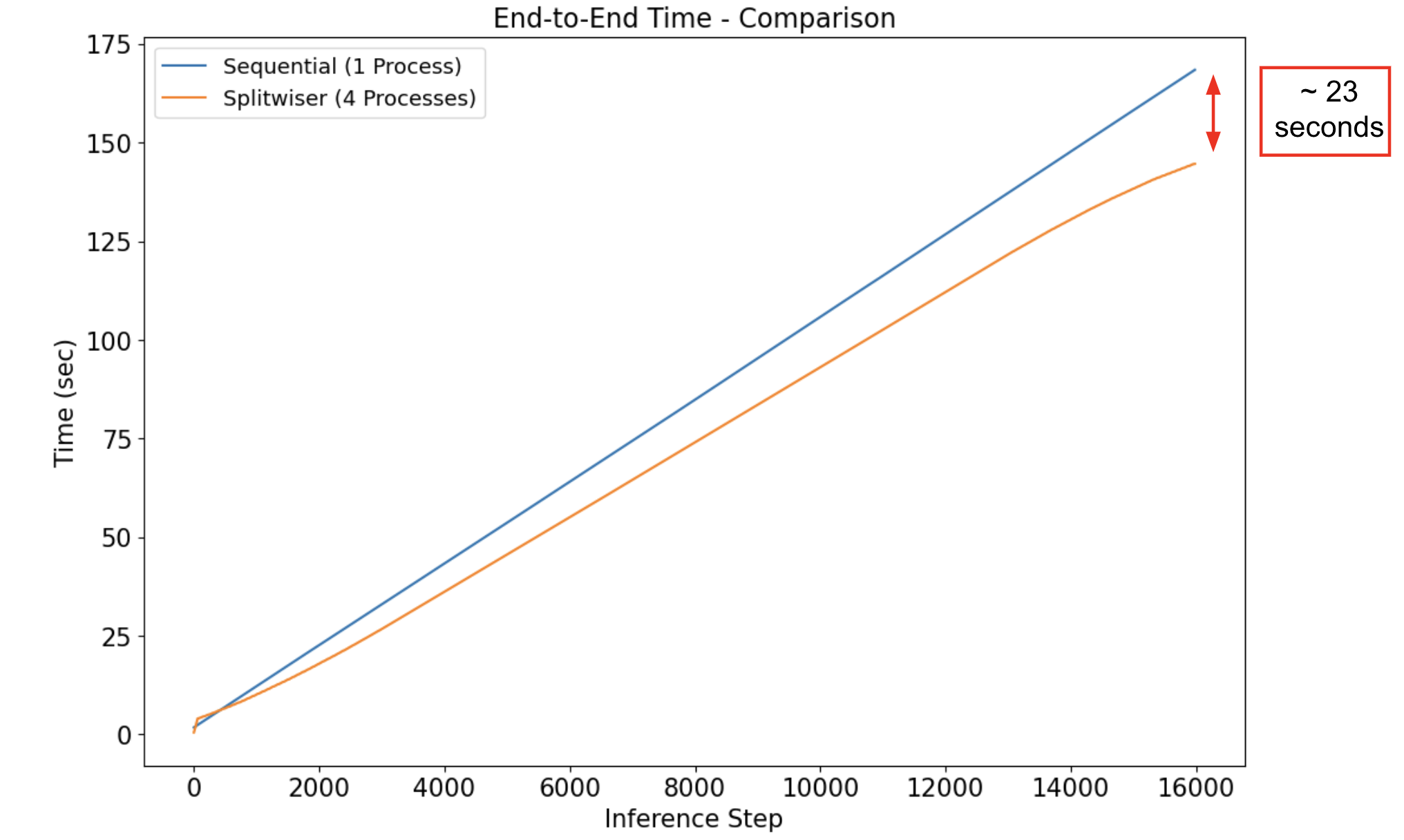}
    \caption{Total elapsed time comparison between huggingface default sequential inference and Splitwiser inference.}
    \label{fig:latency_huggingface}
\end{figure}

\begin{figure}[h]
    \centering
    \includegraphics[width=1\linewidth]{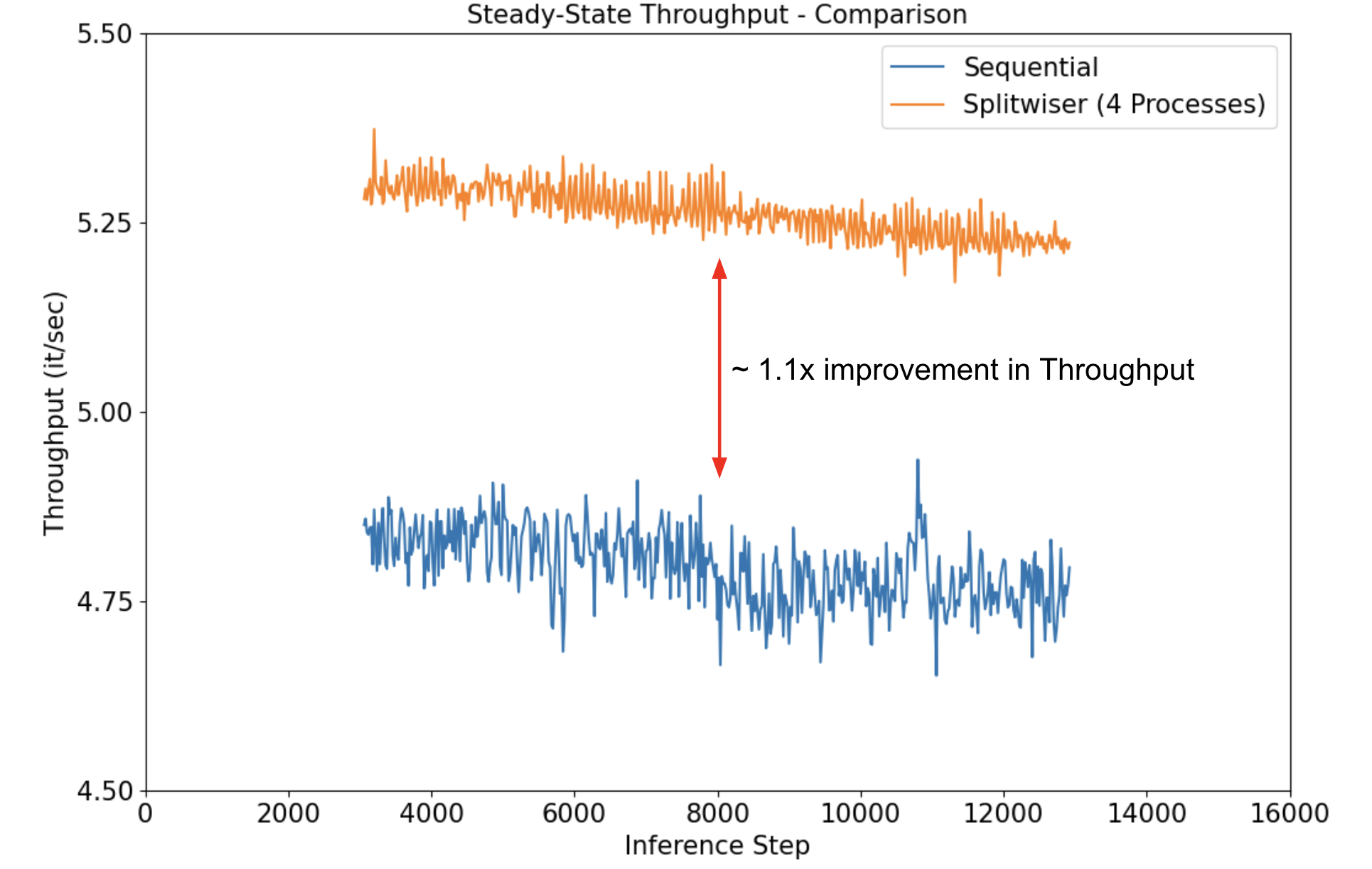}
    \caption{Steady-State throughput comparison between sequential inference and Splitwiser (with 4 processes).}
    \label{fig:throughput_huggingface}
\end{figure}
Based on our pipeline, we outline three independent approaches as described in Section \ref{sec:huggingface}: sequential inference, Splitwiser inference, and Splitwiser+MPS inference. We run multiple experiments to comprehensively evaluate the inference performance across these methods, in terms of latency (end-to-end time) and throughput (iterations run per second). The experimental results are divided as follows:

\begin{enumerate}
    \item \textbf{Latency Analysis}: We run the sequential and Splitwiser implementation over 16,000 inputs, and measure the overall end-to-end time. The result is summarized in Figure \ref{fig:latency_huggingface}, which shows the overall improvement in latency when using Splitwiser inference.

    \begin{figure}[h]
        \centering
        \includegraphics[width=1\linewidth]{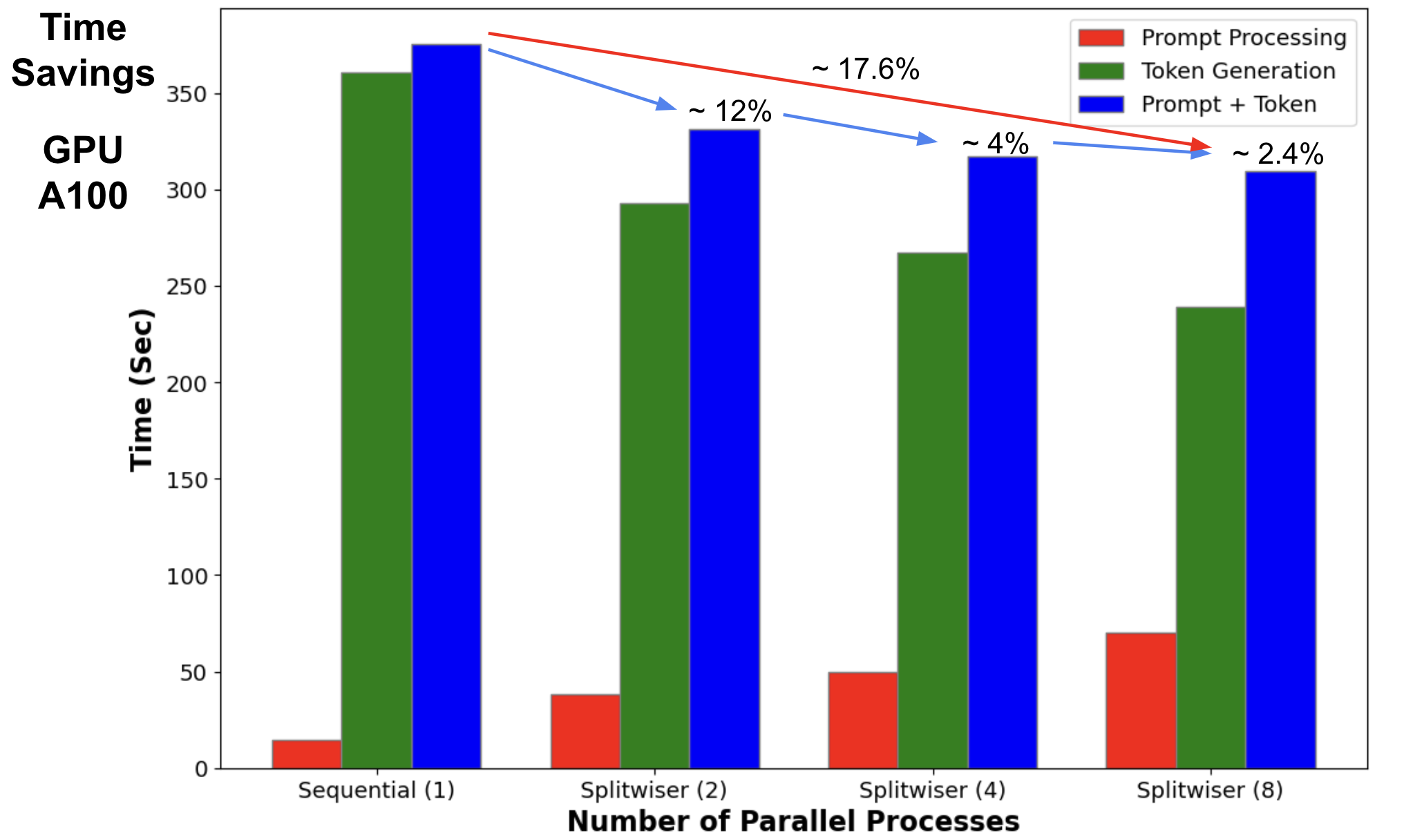}
        \caption{Comparison of end-to-end latency split across the prompt processing phase and token generation phase for Sequential as well as Splitwiser inference. For Splitwiser, we report the latency performance across an increasing number of parallel processes.}
        \label{fig:num_processes_huggingface}
    \end{figure}
    \begin{figure}[h]
        \centering
        \includegraphics[width=1\linewidth]{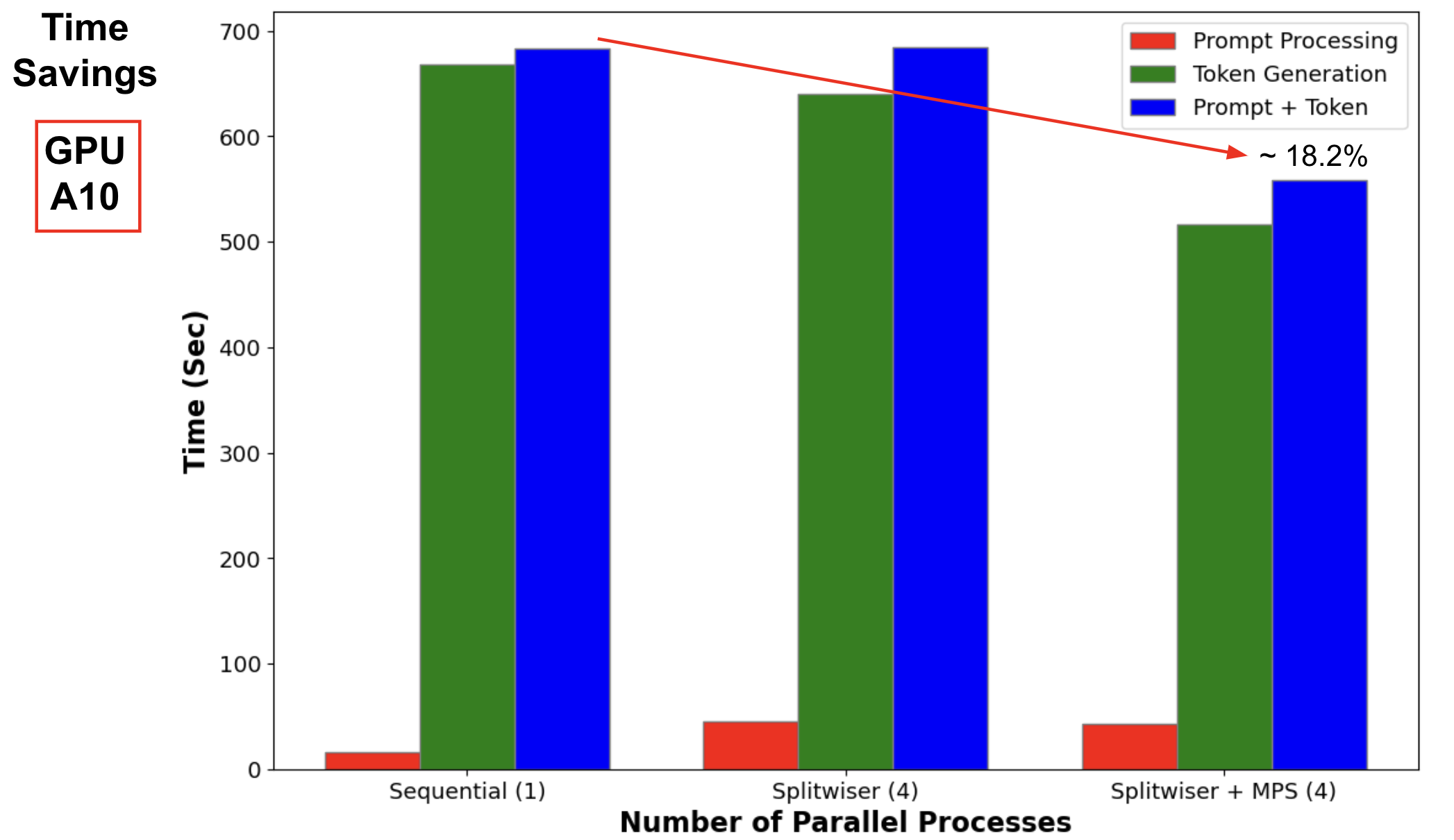}
        \caption{Comparison of end-to-end latency split across the prompt processing phase and token generation phase for Sequential, Splitwiser, and Splitwiser+MPS inference on the A10 GPU.}
        \label{fig:mps_huggingface}
    \end{figure}
    
    \item \textbf{Throughput Analysis}: In this experiment, we compare the steady-state throughput (overall throughput achieved after the initial overhead of setting up multiple processes) of sequential and Splitwiser inference methods. The results are shown in Figure  \ref{fig:throughput_huggingface}, where was can observe a $1.1\times$ improvement in overall throughput by running 4 processes in parallel, compared to a single sequential process.

    \item \textbf{Scaling Number of Parallel Processes}: To further evaluate the impact of running Splitwiser on the overall latency, we run additional experiments where we increase the number of parallel processes on the same GPU. As shown in Figure \ref{fig:num_processes_huggingface}, we can observe that by increasing the number of processes running in parallel, we can achieve faster inference, with the biggest improvement seen between sequential and Splitwiser with 2 parallel processes. Overall, there is a $17.6\%$ reduction in latency when comparing sequential processing with Splitwiser running 8 processes. We can also observe that when increasing the number of parallel processes, the latency for prompt processing goes up. However, the effect of this overhead is negligible compared to the performance speedup introduced by the improved throughput due to parallel processing.

    \item \textbf{Splitwiser + MPS Inference}: In this experiment, we analyze the additional improvement in end-to-end latency by incorporating NVIDIA MPS on top of the Splitwiser multiprocessing design. As shown in Figure \ref{fig:mps_huggingface}, we can observe that on the A10 GPU, we do not see any latency improvement by running Splitwiser alone. However, after combining Splitwiser with MPS, we achieve an overall latency reduction of $18.2\%$.
\end{enumerate}

\subsection{Multiprocessing/MPS with vLLMs}
\begin{figure}[h]
    \centering
    \includegraphics[width=1\linewidth]{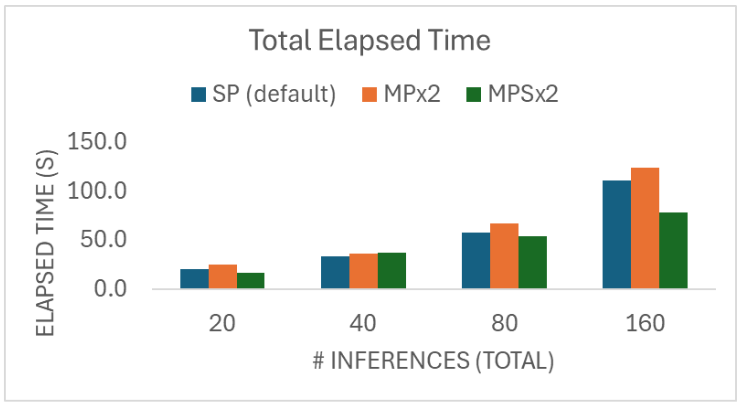}
    \caption{Total elapsed time comparison between default single process (SP), MPx2, and MPSx2.}
    \label{fig:vllm_mp_total_time}
\end{figure}

\begin{figure}[h]
    \centering
    \includegraphics[width=1\linewidth]{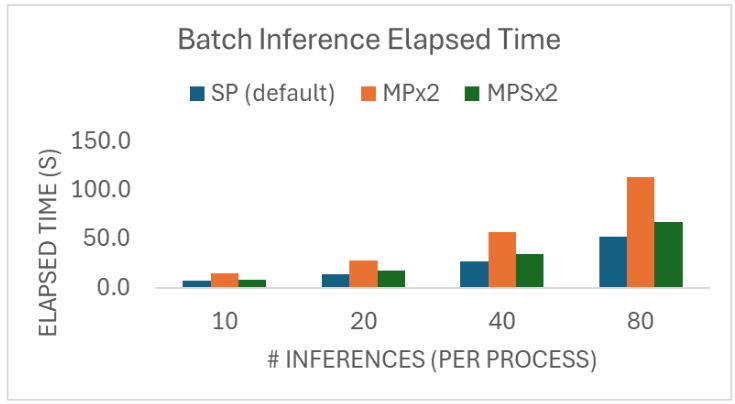}
    \caption{Batch inference elapsed time comparison between default single process (SP), MPx2, and MPSx2. For the MP scenarios, this captures the average batch inference time between the two processes.}
    \label{fig:vllm_mp_batchinf_time}
\end{figure}

\textbf{vLLM + Multiprocessing}
From Figure \ref{fig:vllm_mp_total_time}, we observe a 1.42x speedup completing 160 inference requests comparing a single instance of vLLM (SP) and MPSx2 (two vLLM engines, each processing 80 inference requests). At lower batches, the benefit of MPS throughput is less noticeable as there is an initialization cost to starting multiple processes. Interestingly, we observe that MPx2 without leveraging MPS consistently performs worse than single vLLM. Possibly the throughput gains of multiprocessing are overshadowed by any GPU context switching overhead required switching between the two processes.

A trade-off between throughput and latency is expected and observed when comparing the elapsed time of a batch of inferences in a single process. As shown in Figure \ref{fig:vllm_mp_batchinf_time}, the time taken to process the same batch of inferences in a single process is consistently higher whenever multiprocessing is enabled, regardless of whether MPS is enabled.

\textbf{vLLM Scheduler + Multiprocessing}
While implemented to the extent that inference requests could be serviced successfully, the initial approach was flawed due to the significant overheads and complications from spawning processes on-demand. PyTorch with CUDA tensors necessitate spawning processes instead of forking, and this not only induces heavy initialization cost but also restricts the objects that can be parent-defined as they must be picklable. For prototyping, this required us to remove swapping and CUDA graphs usage as both involved objects that could not be pickled. We discuss improved approaches in the following section.

\section{Conclusion \& Future Considerations}
Overall, we demonstrate that using multiprocessing alone compared to sequential inference can lead to a 17.6\% reduction in latency when running the Huggingface pipeline on the A100. We also observe a $1.1\times$ improvement in steady-state throughput when running 4 processes in parallel using splitwiser, compared to sequential inference. Enabling MPS shows an 18.2\% latency reduction compared to sequential inference when running the Huggingface pipeline on the A10. Finally, on the vLLM pipeline, after enabling MPS, we observe a 1.42x speedup compared to the default single process/single instance execution of vLLM.

From our learnings trying to run multiple processes at the scheduler level, it is understood that starting a process at a lower level is not feasible due to overheads and complications related to process spawning. Therefore, an appropriate next approach would be to instantiate an on-demand prompt process only once and use queues to pass inputs, outputs, and required updates such as memory information. This on-demand prompt process should only be used by the main vLLM scheduler to process simultaneous prompt processing jobs when necessary. However, with this approach, one must take care in handling communication/synchronization overhead. For example, the main vLLM process should continue working asynchronously if communications are still pending from the on-demand process.

Another angle of investigation is to simultaneously execute the pre-processing step of the prompt and token phases. It is noted that the only clear difference between the token and prompt phases is the fetching and processing of initial inputs, either input tokens or KV cache respectively. Therefore, multiprocessing could be reduced to this step or perhaps more investigation can be done to see if these can be completed together without multiprocessing.

\newpage
\bibliographystyle{IEEEtranS}
\bibliography{refs}

\newpage
\section{Appendix}
\label{sec:appendix}
\subsection{Profiling - Batch Size Analysis}
\begin{figure}[!h]
    \centering
    \includegraphics[width=1\linewidth]{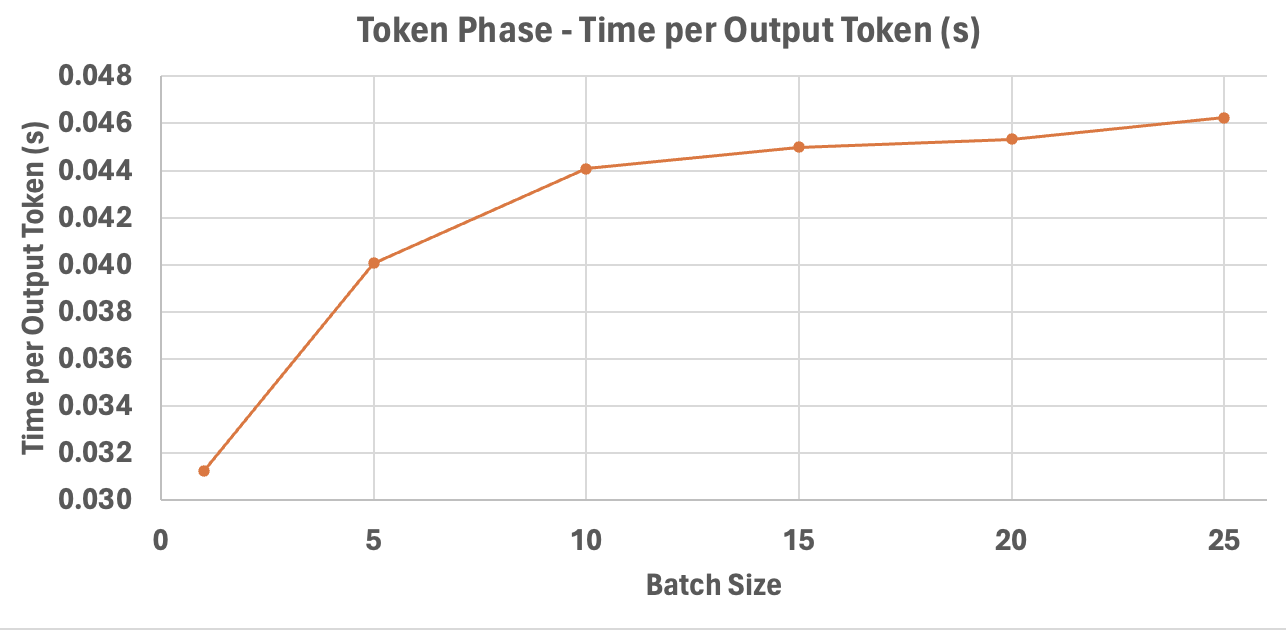}
    \caption{The time per output token in seconds, during the token phase for each batch size.}
\end{figure}
\begin{figure}[h]
    \centering
    \includegraphics[width=1\linewidth]{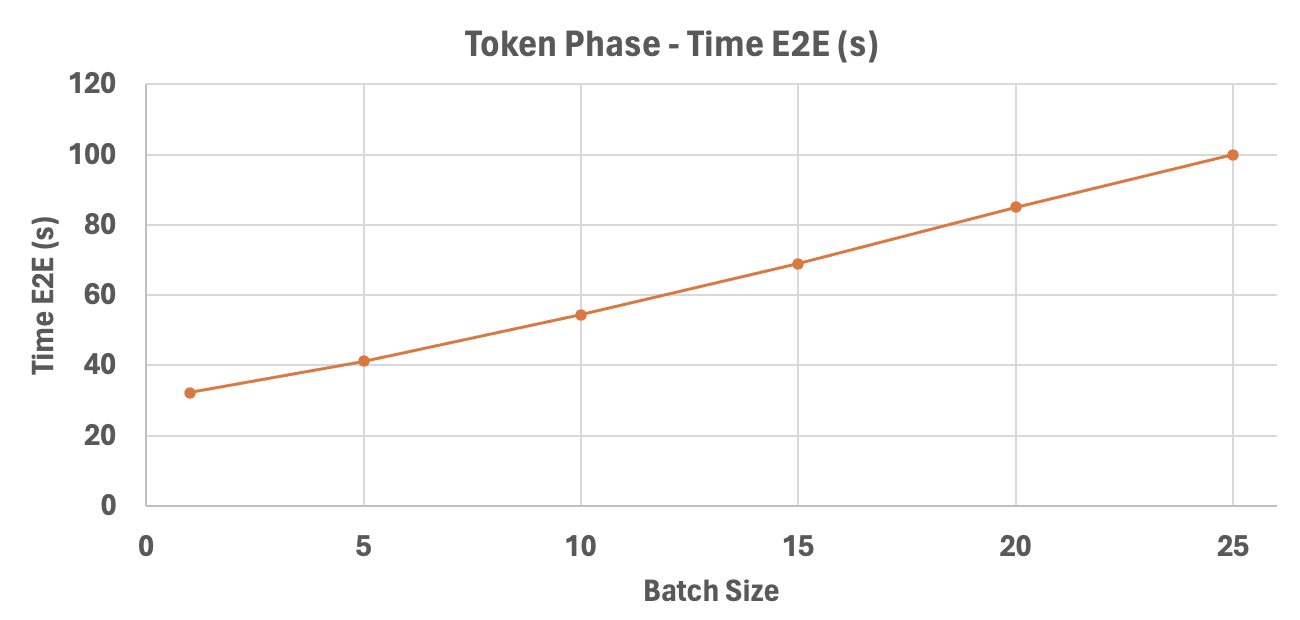}
    \caption{The end-to-end (E2E) time in seconds for each batch size during the token phase.}
\end{figure}
\newpage
\subsection{Profiling - Token Length Analysis}
\begin{figure}[h]
    \centering
    \includegraphics[width=1\linewidth]{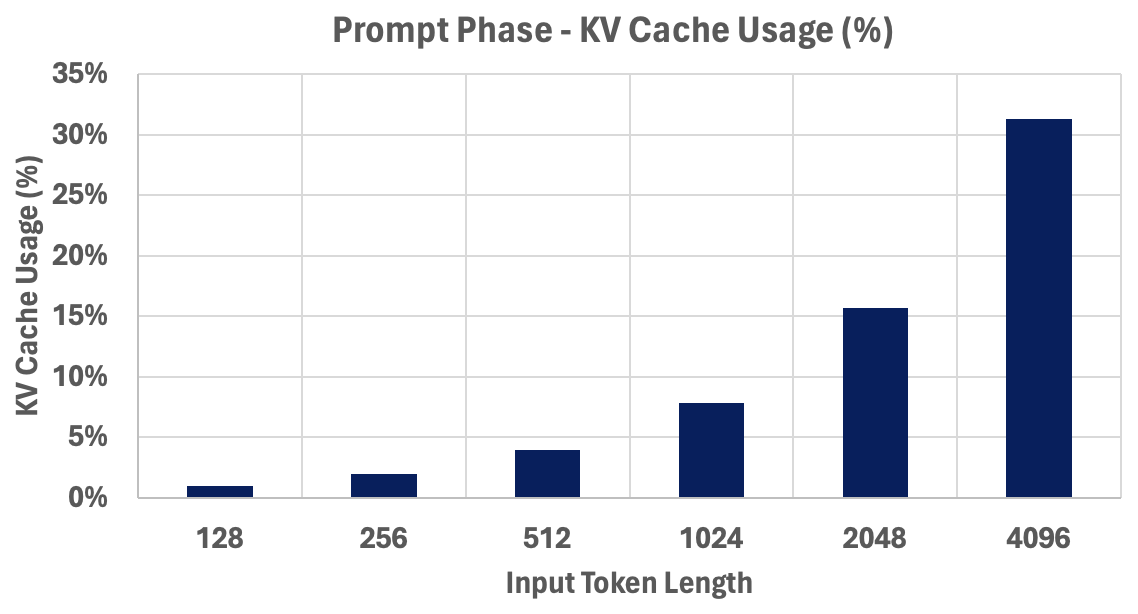}
    \caption{KV-Cache usage (\%) across increasing input token lengths during the prompt phase.}
\end{figure}
\begin{figure}[h]
    \centering
    \includegraphics[width=1\linewidth]{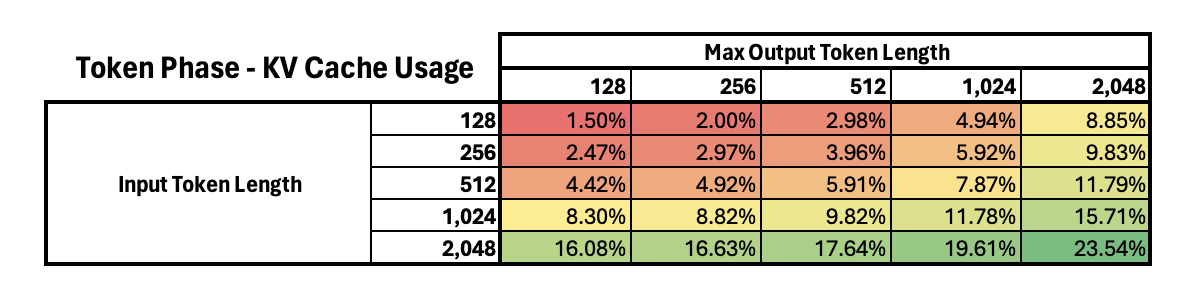}
    \caption{The KV-Cache usage (\%) displayed in a matrix for varying input and maximum output token lengths during the prompt phase.}
\end{figure}

\end{document}